\documentclass[11pt]{article}
\usepackage{a4}
\usepackage{color}
\usepackage{graphicx,graphics}
\usepackage{amsmath,amssymb,amssymb,amstext}
\usepackage{amsthm,amsfonts,amsbsy,pifont,dsfont}
\usepackage{mathrsfs}
\usepackage[english]{babel}
\usepackage[sectionbib]{natbib}
\def\E{\mbox{E}}
\newcommand{\es}{\mbox{\sf \scriptsize{E}}}

\newcommand{\dd}{\mbox{d}}

\def\N{\mbox{N}}

\makeatletter
\newcommand{\periodafter}[1]{#1\hspace{0.4mm}.\hspace{1.3mm}}
\renewcommand\section{\@startsection{section}{1}{0mm}{1.6ex }{0mm}{\noindent\bfseries\periodafter}}
\renewcommand\subsection{\@startsection{subsection}{2}{0mm}{1.6ex }{0mm}{\noindent\bfseries\periodafter}}
\renewcommand{\@seccntformat}[2]{{\csname the#1\endcsname}.\hspace{0.4em}}
\makeatother
\theoremstyle{definition}

\begin{document}

\begin{center}
{{{\Large\sf\bf A probability for classification based on the mixture of Dirichlet process model}}}\\

\vspace{0.5cm}
{\large\sf Ruth Fuentes--Garc\'{i}a$^{*}$, Rams\'es H. Mena$^{**}$ and Stephen G. Walker$^{***}$ }\\
\vspace{0.2cm}
\end{center}
{\sf
\noindent $^*$ Facultad de Ciencias, Universidad Nacional
Aut\'onoma de M\'exico. M\'exico, D.F. 04510, M\'exico.\\
\noindent $^{**}$ \footnote{For correspondence: ramses@sigma.iimas.unam.mx; Tel. +52 (55) 56223543 Ext. 3542; Fax. +52 (55) 56223621 } IIMAS, Universidad Nacional Aut\'onoma de M\'exico. M\'exico, D.F.  04510, M\'exico.\\
\noindent  $^{***}$ University of Kent, Canterbury, Kent, CT2 7NZ, UK.
}

\begin{abstract}
In this paper, we provide an explicit probability distribution for classification purposes. It is derived from the Bayesian nonparametric mixture of Dirichlet process model, but with suitable modifications which remove unsuitable aspects of the classification  based on this model. The resulting approach then more closely resembles  a classical hierarchical grouping rule in that it depends on sums of squares of neighboring values. The proposed probability model for classification relies on a simulation algorithm which will be based on a reversible MCMC algorithm for determining the probabilities, and we provide numerical illustrations comparing with alternative ideas for classification.

\vspace{0.1in} \noindent {\sl Keywords:} Classification; MCMC sampling; MDP model.
\end{abstract}

\section{Introduction}

Suppose we observe data $(y_1,\ldots,y_n)$ which are real numbers on $(-\infty, +\infty)$. The aim is to classify these data into
$k\leq n$ groups and to determine which ones are in the same group.  This is a classic problem and current Bayesian approaches rely on mixture models, such as described in \citet{RichardsonGreen:1997}, or the mixture of Dirichlet process model (see, for example, \citet{Escobar:1994}). In the Richardson and Green model the $k$ is modeled explicitly via
$$p(y|k)=\sum_{j=1}^k w_{j,k}\N(y;\mu_j,\sigma_j^2),$$
where the $w_k=(w_{j,k})_{j=1}^k$ are weights which sum to one. Prior distributions are assigned to $(w_{k},k)$ and $(\mu_j,\sigma_j^2)_{j=1}^\infty$ and inference is made possible via reversible jump MCMC, \citet{Green:1995}.  The likelihood function for $n$ observations is given by
$$l(k,w,\mu,\sigma^2;y,d)\propto \prod_{i=1}^n w_{d_i,k}\,\N(y_i;\mu_{d_i},\sigma_{d_i}^2),$$
where the $(d_i)_{i=1}^n$ are latent variables which pick out the component, less than or equal to $k$, from which the $i$th observation is coming from.

On the other hand, the mixture of Dirichlet process (MDP) model is based on the density function
$$p(y)=\sum_{j=1}^\infty w_j\,\N(y;\mu_j,\sigma_j^2)$$
where the weights $(w_j)_{j=1}^\infty$ sum to one. The parameters $(w_j,\mu_j,\sigma_j^2)$ are assigned distributions and, since the classification ideas we have are based on this model, we will elaborate. So, the $w_1=v_1$ and, for $j>1$,  $w_j=v_j\prod_{l<j}(1-v_l)$ with the $(v_l)$ being independent and identically distributed as $\mbox{beta}(1,\theta)$ random variables for some $\theta>0$.  The $(\mu_j,\lambda_j=\sigma_j^{-2})$ are also independent and identically distributed (the prior) and we consider the prior as
$$\pi(\mu|\lambda)=\N(\mu;0,(c\lambda)^{-1})\,\,\,\,\mbox{and}\,\,\,\,\pi(\lambda)=\mbox{gamma}(\lambda;a,b).$$
In this case the corresponding likelihood function is given by
$$l(w,\mu,\lambda;y,d)\propto \prod_{i=1}^n w_{d_i}\,\N(y_i;\mu_{d_i},\lambda_{d_i}^{-1}).$$
So, in the Richardson and Green model, the $k$ is explicit, but in the MDP model it is implicit, and taken to be the number of distinct $(d_i)$.

However, we are not convinced that either of these models are useful for classification purposes. The key to the problem is that locations of the normal distributions, the $(\mu_j)$, can be arbitrarily close to each other and therefore register as different clusters. So two $\mu_j$ close to each other register as two clusters with a certain probability on the MDP model, since the $d_i$ liking this location may be all one or the other; but does register as two clusters in the Richardson and Green model. Such a scenario may well happen when clusters are not normal based, for example. We would also from this point of view expect the MDP model to perform slightly better than the Richardson and Green model when using as classification modeling. But both methods would over--estimate the number of clusters. We will discuss this issue later in Section 4 when we do some numerical illustrations. Nevertheless, the issue of overestimation of the number of clusters for the MDP has already been known; see, for example, \citet{McGroryTitterington:2007}.

Our approach is not model based yet the starting point is the MDP model; since we believe a classification procedure based on the $(d_i)$ is preferable. Hence, from the MDP model we compute
$p(d|y)$, by integrating out the $(w,\mu,\lambda)$. But this $p(d|y)$ will include many arrangements which are strange for classification purposes. For example, there is positive probability on $d_i$ and $d_{i'}$ both being the same $j$ yet $y_i$ and $y_{i'}$ can be the largest and smallest observation, and observations in between these two extremes are being allocated to different groups.
So, at this point we simply study $p(d|y)$ as a classification probability model and adjust it to eliminate groupings which just don't make any sense. Indeed, it is these such types of $d$ which cause the problems with the MDP model as a classifier in the first place.

So, we first order the $y$, so that $y_1$ is the smallest observation and $y_n$ is the largest observation. We then constrain the $d$ so that the $(d_i)$ are non--decreasing. This ensures that
any group contains only consecutive $y$'s. For example, group 1 would contain a number of the smallest observations; group 2 would contain a number of the next smallest observations; while group $k$ would contain a number of the largest observations. Thus, for any trio of $(y_{i_1}<y_{i_2}<y_{i_3})$, if $y_{i_1}$ and $y_{i_3}$ are in the same group, then so is $y_{i_2}$.
It follows then that our $p^*(d|y)$, with the ordered $y$'s, is given by $p^*(d|y)\propto p(d|y)\,{\bf 1}(d_1\leq \cdots\leq d_n)$. We then show how to sample from $p^*(d|y)$ in order to compute classifications with high probability, and obviously the mode.

In Section 2 we derive and explain our probability model for classification. Section 3 then describes a MCMC algorithm for sampling from this probability model; since for large $n$ the number of possible clusterings is prohibitively large to compute directly. Section 4 then presents numerical illustrations based on a toy example of 10 data points whereby all probabilities can be computed and the well known and widely studied galaxy data set.

\section{The classification probability model}

Given the outline in the Introduction, our first task is to compute $p(d|y)$ based on the MDP model. Now
\[p(d,y|\mu,\lambda,w)=\prod_{i=1}^n w_{d_i}\,\N(y_i;\mu_{d_i},\lambda_{d_i}^{-1})\]
and so
\[\begin{array}{ll}
p(d,y) & =\E\left\{\prod_{i=1}^n v_{d_i}\prod_{l<{d_i}}(1-v_l)\right\}\,\,\prod_{j=1}^\infty \int \prod_{d_i=j}\,\N(y_i;\mu,\lambda^{-1})\,\pi(\dd\mu,\dd\lambda) \\ \\
& = \prod_{j=1}^\infty \left\{\theta\,\int v^{n_j}(1-v)^{m_j+\theta-1}\,\dd v\right\}\,\left\{\int \prod_{d_i=j}\,\N(y_i;\mu,\lambda^{-1})\,\pi(\dd\mu,\dd\lambda)\right\}.
\end{array}\]
Here, $n_j=\sum_{i=1}^n {\bf 1}(d_i=j)$ and $m_j=\sum_{i=1}^n {\bf 1}(d_i>j)$.
The first term in the product is given by
\[\prod_{j=1}^\infty \frac{\theta\Gamma(1+n_j)\Gamma(\theta+m_j)}{\Gamma(1+\theta+n_j+m_j)}\]
and the second term is easily found to be given by
\[\prod_{j=1}^\infty \frac{\Gamma(a+n_j/2)b^a\sqrt{c}}{\left\{b+S_j^2/2\right\}^{a+n_j/2}\,\sqrt{c+n_j}\Gamma(a)},\]
where
\[S_j^2=\sum_{d_i=j}y_i^2-\frac{n_j\bar{y}_j^2}{1+c/n_j}\] and
\[\bar{y}_j=n_j^{-1}\sum_{d_i=j} y_i.\]
Hence,
\[p(d|y)\propto \prod_{j=1}^\infty \frac{\theta\Gamma(1+n_j)\Gamma(\theta+m_j)}{\Gamma(1+\theta+n_j+m_j)}\,\frac{\Gamma(a+n_j/2)b^a\sqrt{c}}{\left\{b+S_j^2/2\right\}^{a+n_j/2}\,\sqrt{c+n_j}\Gamma(a)}.\]
This then is the probability of classification based on the MDP model.

At this point, we simply focus on $p(d|y)$ and assess it as a probability model for classification. So, without loss of generality, we take the $y$'s to be ordered, with $y_1$ being the smallest observation and $y_n$ being the largest. For reasons then given in the Introduction, we would now for classification purposes only wish to consider the $(d_i)$ to be non--decreasing. Hence, we consider
$p^*(d|y)\propto {\bf 1}(d_1\leq \cdots \leq d_n)\,p(d|y)$. We also impose the constraint that if there are $k$ distinct $(d_i)$ then $d_n=k$.

Our observation now is that $d$ is completely determined by $(k,n_1,\ldots,n_k)$ whereby $k$ is the number of distinct $(d_i)$ and $n_j$ is the number of the $d_i$ equal to $j$. Hence,
\[p(k,n_1,\ldots,n_k)=\kappa\,\prod_{j=1}^k \frac{\theta\Gamma(1+n_j)\Gamma(\theta+m_j)}{\Gamma(1+\theta+n_j+m_j)}\,\frac{\Gamma(a+n_j/2)b^a\sqrt{c}}{\left\{b+S_j^2/2\right\}^{a+n_j/2}\,\sqrt{c+n_j}\Gamma(a)}\]
where $\kappa$ is the normalizing constant, and now we define $m_j=n-n_1-\cdots-n_j$ and
\[S_j^2=\sum_{i=n_{j-1}^*+1}^{n_j^*}y_i^2-\frac{n_j\bar{y}_j^2}{1+c/n_j}\]
and
\[\bar{y}_j=n_{j}^{-1}\sum_{i=n_{j-1}^*+1}^{n_j^*}y_i,\]
with $n_j^*=n_1+\cdots+n_j$ and $n_0^*=0$. Note that, for a given sample size $n$, the support of this probability runs over the set of compositions of the integer $n$ rather than on the number of partitions of a set with $n$ elements typically found in the MDP or other exchangeable partition probability functions settings encounter in the Bayesian nonparametric literature.

This probability model for classification  is a highly suitable  and necessary adaption of the probability model for classification based on the MDP model. It can be seen to depend fundamentally on the sample variances of the observations in the same group. So the lower the sample variances, the higher the probability. The rule of having $k=n$ groups is countered by the probability being a product of $k$ terms. The probability depends on the parameters $(\theta,a,b,c)$, which would basically have the same interpretation as if we were using a MDP model for the data. So, for example, if $\theta$ is big, which implies a large number of groups in the MDP model, its role can be seen explicitly in $p(k,n_1,\ldots,n_k)$, since we would have the term $\theta^k$ and so encourages large $k$.

We also note that attempts have been made to emphasize the suitable $(k,n_1,\ldots,n_k)$ by using alternative nonparametric mixing  prior distributions to the Dirichlet process, which constitutes the MDP model, and which put more weight on configurations which are realistic, see \citet{LijoiMenaPrunster:2007}. However, positive mass is still being put on ``ridiculous" configurations which will lead to overestimation of $k$. Our approach, in light of this, is remarkably obvious in that we put zero weight on all but realistic configurations.

Here we also mention the problem of what happens if a new piece of data arrives. Our approach is not to assume a clustering for the existing data has been set and we decide into which group, possibly a new one, the extra piece of data should be put; but rather we merely recompute $p(k,n_1,\ldots,n_k)$ with all the data, including the additional piece. We do not see any other approach as being relevant here.

We will compare our approach with a routine in the package \emph{R}, a hierarchical clustering routine based on local sums of squares, so in principle is not unlike the idea of working with sample variances. The routine is labeled \texttt{hclust} in \emph{R} and is based on an original algorithm appearing in \citet{Ward:1963}.

\section{Sampling the model}\label{sec:Sampling}

The basic idea for sampling from $p(k,n_1,\ldots,n_k)$ will be a split--merge MCMC algorithm. So at each iteration one of 2 types of move will be proposed: a split, whereby a group of size bigger than 1 is divided into 2 groups so $k$ is increased by 1; and a merge, whereby 2 groups are combined into 1 group so $k$ is decreased by 1.
The idea for sampling from $p(k,n_1,\ldots,n_k)$ can be seen as a reversible jump MCMC algorithm, and for ease of exposition we will describe the algorithm using latent variables and the specification of a joint density for a configuration conditional on a $k$: so let $n^{(j)}$ for $j=1,\ldots,n$ be a clustering for $j$ groups, and consider
\[p(k,n^{(1)},\ldots,n^{(n)})  = p(k,n_1,\ldots,n_k) \! \prod_{j=k+1}^n p(n^{(j)}|n^{(j-1)}) \,\prod_{j=1}^{k-1} p(n^{(j)}|n^{(j+1)}),\]
which is based on  a recent idea described in \citet{Walker:2009}.  The concern is that the marginal density for $(k,n_1,\ldots,n_k)$ is unchanged; which is the case, as is evidently obvious. Now given a $k$ and $n^{(k)}$ we propose a move to $k+1$ with probability 1/2 and to $k-1$ with probability 1/2 (with obvious modifications if $k=1$ or $k=n$). We need to therefore sample $n^{(k+1)}$ from $p(n^{(k+1)}|k,n^{(k)})$ and $n^{(k-1)}$ from $p(n^{(k-1)}|k,n^{(k)})$. The former is achieved by finding an existing group with size $>1$,  and then we split this group into 2. If uniform distributions are used for both operations then
\[p(n^{(k+1)}|k,n^{(k)})=\frac{1}{n_g^{(k)}(n_s^{(k)}-1)},\]
where $n_g^{(k)}$ is the number of groups of size $>1$, from $(n_1,\ldots,n_k)$, and $n_s^{(k)}$ is the size of this group chosen.
The latter is obtained by merging two neighboring groups and so with a uniform distribution we have
\[p(n^{(k-1)}|k,n^{(k)})=1/(k-1).\]
Therefore, the sampler carries out each step through a Metropolis-Hastings scheme. When at state $x^{(k)}=(k,n_1,\ldots,n_k)$ the acceptance probability to move to state $x^{(k+1)}=(k+1,n_1,\ldots,n_{k+1})$ is
\begin{eqnarray*}
\alpha(x^{(k)},x^{(k+1)})= \min \left \{ 1, \frac{1-q}{q} \frac{p(x^{(k+1)})}{p(x^{(k)})} \frac{n_{g}^{(k)} (n_{s}^{(k)}-1)}{k} \right \},
\end{eqnarray*}
where $q=1/2$. On the other hand, if instead two groups, $(n_{s1}^{(k)},n_{s2}^{(k)})$, in $x^{(k)}$ are selected and we  attempt to merge them, then the  acceptance probability for this move is
\begin{eqnarray*}
\alpha(x^{(k)},x^{(k-1)})= \min \left \{ 1, \frac{q}{1-q} \frac{p(x^{(k-1)})}{p(x^{(k)})} \frac{k-1}{(n_{s1}^{(k)}+ n_{s2}^{(k)}-1)n_g^{(k-1)}} \right \},
\end{eqnarray*}
where $n_{g}^{(k-1)}$ is the cardinality of the set containing all groups with more than one observation in $x^{(k-1)}$.

To improve the algorithm we then shuffle the $n^{(k)}$ by selecting adjacent groups, $(n_{s1},n_{s2})$ and attempting to change them into  $(n_{s1}^*,n_{s2}^*)$ in such a way that both $n_{s1}^*$ and $n_{s2}^* \geq 1$. The shuffle is based on the idea of putting the two groups together and then uniformly splitting into 2 groups. The acceptance probability is then given by
\begin{eqnarray*}
\alpha(x,x^*)= \min \left \{ 1, \frac{p(x^*)}{p(x)}\frac{(n_{s1}^*+ n_{s2}^*-1)}{(n_{s1}+ n_{s2}-1)}
\right \}.
\end{eqnarray*}
These acceptance probabilities all follow from the expression $p(k,n^{(1)},\ldots,n^{(n)})$ and the cancelations which occur when evaluating the  ratios of neighboring $k$.

The algorithm is effectively a joint Metropolis--Hastings and Gibbs algorithm, rather than a reversible jump MCMC algorithm. The dimension is fixed and so no special considerations arise on this issue. The only necessary consideration is that if $p(n^{(k)}|n^{(k-1)})>0$ then $p(n^{(k-1)}|n^{(k)})>0$, and vice versa.  Neither, if one is even needed, have we had to worry about a Jacobian, since we are not basing the moves on transformations of variables between different dimensions. We believe it is more explicit to understand reversible jump MCMC from this perspective.

Here we consider a more general idea for sampling, based on the notion of a joint density
$$p(k,n^{(1)},\ldots,n^{(n)})  = p(k,n^{(k)})\,p(n^{(2)},\ldots,n^{(k-1)},n^{(k+1)},\ldots,n^{(n-1)}|n^{(k)}).$$
Then it is easy to see how the reversible jump MCMC arises from this model. But we can seek alternative, and more general strategies, and one such is based on the idea of
$$p(n^{(2)},\ldots,n^{(k-1)},n^{(k+1)},\ldots,n^{(n-1)}|n^{(k)})=p(n^{(2)})\,\prod_{j=3}^{k-1}p(n^{(j)}|n^{(j-1)})\,\,\prod_{j=k+1}^{n-1}p(n^{(j)}|n^{(j-1)}).$$
where $p(n^{(k)}|n^{(k-1)})$ is the probability density for a split move described earlier, and $p(n^{(2)})$ is the correct density for $n^{(2)}$ given $k=2$, and is easy to sample since
$n^{(2)}$ can be represented by a single number between $1$ and $n-1$.
Then it is easy to see that a move from $x^{(k)}$ to $x^{(k')}$, with $k'\in(k-1,k+1)$ can be achieved by first sampling $x^{(k+1)}$ from $p(n^{(k+1)}|n^{(k)})$
and $x^{(k-1)}$ from the density $p(n^{(2)})\,\prod_{j=3}^{k-1}p(n^{(j)}|n^{(j-1)})$, which is done by sampling $n^{(2)}$, then $n^{(3)}$, and so on, up to $n^{(k-1)}$.
If $p(n^{(k)}|n^{(k-1)})=0$ then the proposed move is rejected and $(k,n^{(k)})$ is kept. On the other hand, if $p(n^{(k)}|n^{(k-1)})>0$ then
a move to $k+1$, proposed with probability $1/2$ is accepted with probability
$$\min\left\{1,\frac{p(k+1,n^{(k+1)})\,p(n^{(k)}|n^{(k-1)}) }{p(k,n^{(k)})\,p(n^{(k+1)}|n^{(k)})}\right\},$$
or else a move to $k-1$ is proposed and  is accepted with probability
$$\min\left\{1,\frac{p(k-1,n^{(k-1)})\,p(n^{(k)}|n^{(k-1)}) }{p(k,n^{(k)})\,p(n^{(k-1)}|n^{(k-2)})}\right\}.$$
While in this particular case we do not claim an improvement using this alternative, the point is that there are alternatives to be considered.
In this way it can be seen that the reversible jump MCMC methodology can be viewed as  a special case of a particular idea formulated by the notion of a joint density
$$p(k,n^{(1)},\ldots,n^{(n)})  = p(k,n^{(k)})\,p(n^{(2)},\ldots,n^{(k-1)},n^{(k+1)},\ldots,n^{(n-1)}|n^{(k)}).$$
To see how the algorithms work here; let us ease the notation by writing $\theta_k=n^{(k)}$ and $\theta_{-k}$ be the $\{n^{(j)}; j\ne k\}$. Then, at $(k,\theta_k)$, we sample $\theta_{-k}$ from the full conditional, but in the original algorithm only need $\theta_{k-1}$ and $\theta_{k+1}$, and then do a Metropolis--Hastings step for $(k,\theta_k)$ where the proposal is to
complete the joint density with $k+1$, or $k-1$,  and the retention of $\theta_k$.

\section{Numerical illustrations}

In order to underline the kind of results that can be obtained by our approach we first consider a small data set; small enough $(n=10)$ so that we can provide exact computations of
probabilities for all $(k,n_1,\ldots,n_k)$. We then illustrate our approach with a real data set; the galaxy data set.

\subsection{Small data set}

\begin{center}
\begin{figure}[!htbp]
\begin{center}
\includegraphics[scale=1]{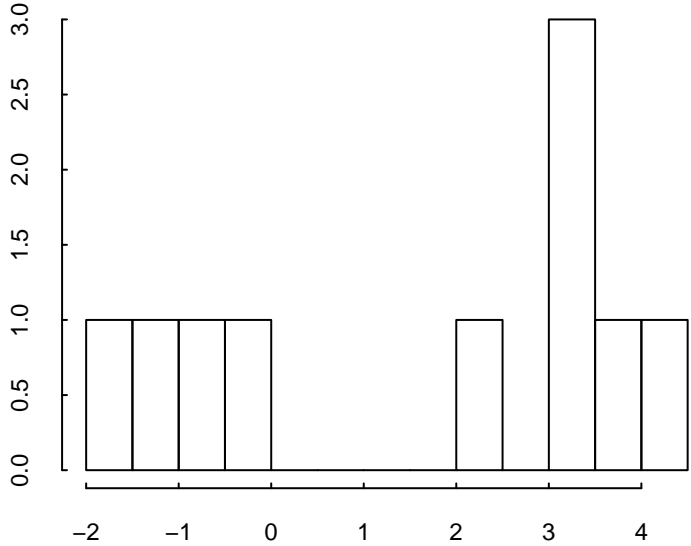}
\caption{Histogram for small data set.}\label{fig:small}
\end{center}
\end{figure}
\end{center}

Suppose the set of ordered observations is $y=($-1.522, -1.292, -0.856, -0.104, 2.388, 3.080, 3.313, 3.415, 3.922, 4.194$)$, and a histogram of the data is shown in Figure~\ref{fig:small}, from this it is evident that 2 groups are the most likely option.
Table~\ref{tab:toy} gives the probabilities of having $k$ groups using the MDP approach, and our proposed approach. In
both cases we use the prior specification of  parameters as $\theta=a=b=1$ and $c=0.1$. For the MDP model we
computed the exact probabilities for each of the 115,975 possible
partitions of $y$, and then using them to obtain the exact posterior probabilities for each
$k\in\{1,\ldots,10\}$. The highest probability in this case is allocated to
$k=3$. Further inspection among the partitions indicates that the highest
posterior probability of $0.332$ corresponds to the classification involving 2 groups;
$\left\{[y_1,y_2,y_3,y_4],[y_5,\ldots,y_{10}]\right\}$, which corresponds to $(n_1,n_2)=(4,6)$.

On the other hand, the exact probabilities, $p^*(k)$,
computed over all the 512 possible configurations, assign the highest
probability to $k=2$. As with the MDP case, the classification with the highest
probability corresponds to
$\left\{[y_1,y_2,y_3,y_4],[y_5,\ldots,y_{10}]\right\}$; but in this case
with the considerably higher probability of  $0.833$. Clearly, considering the order
of the $y$ limits the support considerably, from set partitions to integer
compositions, withdrawing all inadequate partitions for classification
purposes and hence leading to an improved estimator for the
number of groups.

The last four columns in Table~\ref{tab:toy} show the estimates of $p^*(k)$
based on the MCMC schemes described in Section~\ref{sec:Sampling}, with $10,000$ and $100,000$ iterations following
a burn in period of $1,000$ and $10,000$ iterations respectively. The results are matching the exact
probabilities and from these it is evident that both schemes are valid, although the first appears to converge faster than the second.

\begin{center}
\begin{figure}[!htbp]
\begin{center}
\includegraphics[scale=0.5]{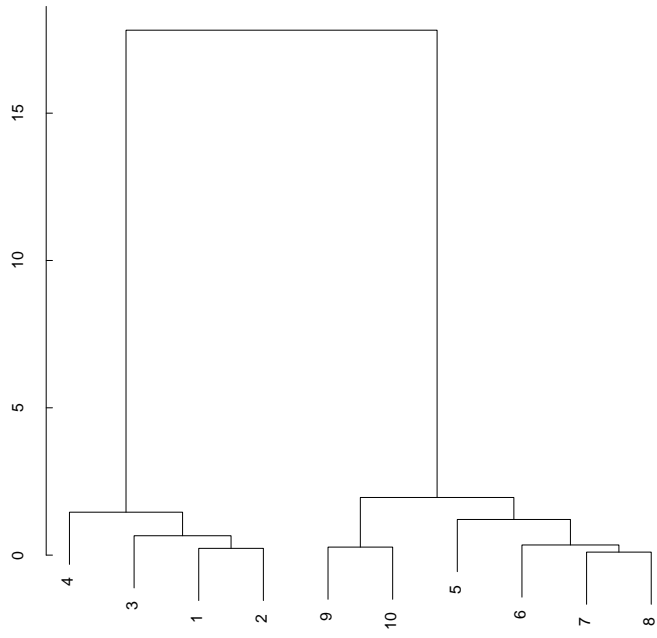}
\caption{Dendogram for the small data set example..}\label{fig:dendo}
\end{center}
\end{figure}
\end{center}

Our results  are in agreement with the hierarchical agglomerative
clustering, using Ward's (1963) approach, which reduces the number of groups from
$k$ to $k-1$ by minimizing the local sum of squares; see
Figure~\ref{fig:dendo}. It is obvious from this that 2 groups are by far and away the preferred choice.

\subsection{Galaxy data set}

Here we consider the galaxy data set, studied in \citet{Roeder:1990}. It is widely used in the literature to illustrate methodology for
mixture modeling.  In this case the sample size is $n=82$ and so we would need to compute $2^{81}$ probabilities to obtain all the possible configurations.

Therefore, we will use the first MCMC algorithm proposed in Section~3 to obtain
the probabilities. We undertake this approach, with the same
parameter specifications as in the above small
data set example and 10000 iterations after 1000 of burn in period. The MCMC estimates result in $p^*(k=3)=0.997$ and
$p^*(k=4)=0.003$, with the highest probability of 0.677 on the configuration
$(n_1,n_2,n_3)=(7,72,3)$. The same results are attained with the second scheme of Section~\ref{sec:Sampling} but with a higher number of simulations.

Depending on the parameter values, e.g. the total mass parameter
$\theta$; the Bayesian nonparametric mixture model favors between 5
and 6 groups; see for example \citet{EscobarWest:1995} and \citet{LijoiMenaPrunster:2005}. Similar results are achieved in the finite mixture setting as
in \citet{RichardsonGreen:1997}. All of these approaches seem to be overestimating the number of
groups, as noted from results reported in \citet{McGroryTitterington:2007}.


\section*{Acknowledgements}
The first author gratefully acknowledges the Mexican Mathematical Society and the Sofia Kovalevskaia Fund, and the second author gratefully acknowledges CONACYT for Grant No. J50160-F, for allowing them to travel to UK, where the work was completed during a visit to the University of Kent.

\renewcommand\refname
\newpage

\begin{center}
\begin{table}
\begin{center}
\begin{tabular}{|l||l||l||c|c||c|c|}
   \hline

          &       &  Exact    & M1 (A) & M1 (B)  &M2 (A)& M2 (B)\\
   { $k$} & {MDP} &  $p^*(k)$ & ${\widehat p^*}(k)$&${\widehat p^*}(k)$ &${\widehat p^*}(k)$&${\widehat p^*}(k)$\\

   \hline
   \hline
   1 &  $0.00619$           & $0.04535$           &  $0.04760$  &  $0.04709$  & $0.04450$  & $0.04794$     \\
   2 &  $0.37634$           & $0.88622$           &  $0.88480$  &  $0.88375$  & $0.84770$  & $0.88376$     \\
   3 &  $0.39729$           & $0.06597$           &  $0.06710$  &  $0.06652$  & $0.10130$  & $0.06482$     \\
   4 &  $0.17298$           & $0.00240$           &  $0.00050$  &  $0.00250$  & $0.00650$  & $0.00348$     \\
   5 &  $0.04088$           & $0.00006$           &  --         &  $0.00011$  &  --        &  --           \\
   6 &  $0.00578$           & $1.00\,\es{-6}$     &  --         &  $0.00003$  &  --        &  --           \\
   7 &  $0.00051$           & $1.31\,\es{-8}$     &  --         &    --       &   --       &   --          \\
   8 &  $0.00003$           & $1.22\,\es{-10}$    &  --         &    --       &   --       &   --          \\
   9 &  $8.38\,\es{-7}$     & $7.44\,\es{-13}$    &  --         &    --       &   --       &   --          \\
   10 & $1.12\,\es{-8}$     & $2.26\,\es{-14}$    &  --         &    --       &   --       &   --          \\
   \hline
\end{tabular}
\end{center}
\caption{ \label{tab:toy} Probabilities on the different number of groups for the small data set example. The MDP results correspond to exact posterior probabilities. The probabilities $p^*(k)$ and ${\widehat p^*}(k)$ for the classification model correspond to the exact and MCMC estimates, respectively. The columns labeled M1 and M2 refer to the two sampling schemes described in Section~\ref{sec:Sampling} with (A)  10000 iterations after a 1000 burn in period  and (B) 100000 iterations after a 10000 burn in period.}
\end{table}
\end{center}

\end{document}